\documentclass[runningheads]{llncs}
\usepackage[T1]{fontenc}
\usepackage{amsmath}
\usepackage{amssymb}
\usepackage{booktabs}

\usepackage{graphicx}

\begin{document}

\title{Brain-Aware Readout Layers in GNNs: Advancing Alzheimer's early Detection and Neuroimaging }

\author{Jiwon Youn\inst{1} \and
Dong Woo Kang\inst{2} \and
Hyun Kook Lim\inst{3} \and
Mansu Kim\inst{1,*}}

\authorrunning{J. Youn et al.}

\institute{AI Graduate School, Gwangju Institute of Science and Technology, Gwangju, Republic of Korea
 \and
Department of Psychiatry, Seoul St. Mary’s Hospital, College of Medicine, The Catholic University of Korea, Seoul, Republic of Korea
\and
Department of Psychiatry, Yeouido St. Mary’s Hospital, College of Medicine, The Catholic University of Korea, Seoul, Republic of Korea\\
\email{mansu.kim@gist.ac.kr}}

\maketitle             

\begin{abstract}
Alzheimer's disease (AD) is a neurodegenerative disorder characterized by progressive memory and cognitive decline, affecting millions worldwide. Diagnosing AD is challenging due to its heterogeneous nature and variable progression. This study introduces a novel brain-aware readout layer (BA readout layer) for Graph Neural Networks (GNNs), designed to improve interpretability and predictive accuracy in neuroimaging for early AD diagnosis. By clustering brain regions based on functional connectivity and node embedding, this layer improves the GNN's capability to capture complex brain network characteristics. We analyzed neuroimaging data from 383 participants, including both cognitively normal and preclinical AD individuals, using T1-weighted MRI, resting-state fMRI, and FBB-PET to construct brain graphs. Our results show that GNNs with the BA readout layer significantly outperform traditional models in predicting the Preclinical Alzheimer's Cognitive Composite (PACC) score, demonstrating higher robustness and stability. The adaptive BA readout layer also offers enhanced interpretability by highlighting task-specific brain regions critical to cognitive functions impacted by AD. These findings suggest that our approach provides a valuable tool for the early diagnosis and analysis of Alzheimer's disease.

\end{abstract}

\section{Introduction}
\noindent Alzheimer's disease (AD) is characterized by progressive impairment of memory and cognitive functions, affecting approximately 47 million people worldwide. A major challenge in diagnosing AD is its heterogeneity among individuals, as it typically manifests as a decline in cognitive abilities. Although the onset of AD correlates with the accumulation of amyloid proteins beyond a certain threshold, these proteins are not the direct cause of AD, and the future progression patterns also vary among individuals \cite{silverman1999clinical,serrano2021apoe,selkoe2016amyloid}.

Early diagnosis of AD is critical due to no effective treatments for dementia associated with AD \cite{ael2003early}. Recently, it has been reported that a preclinical stage of AD is asymptomatic and characterized by extremely minor cognitive declines, which can be captured by the Preclinical Alzheimer's Cognitive Composite (PACC) score \cite{hanseeuw2019association}. The PACC score provides a measure of cognitive abilities that may decline even before noticeable symptoms appear by capturing subtle changes in episodic memory, executive function, and attention \cite{donohue2014preclinical}.

For decades, the ATN framework, $\beta$-amyloid deposition (A), tau pathology (T), and neurodegeneration (N) play important roles in understanding AD, has been focused on understanding AD pathology and identifying related biomarkers \cite{jack2018nia}. Recent studies have successfully demonstrated to separate AD and cognitive normal(CN) based on T1-weighted magnetic resonance imaging (MRI), while having a limited performance on early diagnosis separating CN with mild cognitive impairment (MCI) \cite{salvatore2015magnetic}. Another study based on positron emission tomography (PET) reported that reported significant reductions in metabolic activity in key brain regions associated with Alzheimer's disease, while encountering challenges related to resolution issues \cite{hoilund2023fdg,chapleau2022role}. Recent studies have highlighted disruptions in the default mode network through functional MRI, showing connectivity reductions that correlate with cognitive declines and predict progression from mild cognitive impairment to Alzheimer's disease \cite{sheline2013resting}.

Recent Graph Neural Networks (GNNs) studies have demonstrated high effectiveness across various tasks, including recommendation systems, computer vision, and neuroimaging \cite{wu2020comprehensive}. Specifically, recent bioinformatics studies have demonstrated that GNNs achieve higher performance than traditional machine learning techniques in tasks such as disease prediction, drug discovery, and patient classification  \cite{boll2024graph,zhang2023graph,zhou2020graph,jiang2021could}. Moreover, neuroimaging studies have demonstrated that GNNs are highly effective at capturing the intricate connectivity and patterns within brain networks outperforming traditional analysis techniques \cite{li2021braingnn}. These studies suggest that GNN might have the potential to understate complex brain networks due to its ability to capture complex relationships within data.

In this paper, we propose a novel brain-aware readout layer (BA readout layer) for GNNs to improve their interpretability. By applying the proposed readout layer, the GNN model can not only capture the complex patterns inherent in brain data but also identify groups of regions whose functionalities are closely related to each other. Our major contribution is as follows:  1) The proposed BA readout layer is more interpretable than the conventional readout layer by leveraging brain functional network. 2) We demonstrated the flexibility of the proposed BA readout layer by implementing it across different GNN frameworks. 3) The empirical experiment on real data has been performed to demonstrate the effectiveness of the proposed model compared with competing methods.

\section {Related Works}
GNNs are a powerful framework for learning from graph-structured data. Unlike traditional neural networks that excel at processing grid-like data such as images or sequences, GNNs are designed to handle data represented in the form of graphs. In a graph $G=(\mathcal{V},\mathbf{E})$, the set $\mathcal{V}$ represents nodes (or vertices) and $\mathbf{E}$ represents edges (or relationships) between the nodes.

The fundamental concept of GNN is the iterative updating of node representations by aggregating information from neighboring nodes in the graph using an aggregation layer. This process enables GNN to capture complex relational dependencies and structural properties inherent in the data. Finally, the readout layer has been used to aggregate node representations to produce a single representation for the entire graph.

\subsection{Aggregation Layer}
\subsubsection{Graph Convolutional Networks(GCN).}  GCN is an aggregation layer based on convolution operation for graphs by using a localized first-order approximation of spectral graph convolutions \cite{kipf2016semi}. The fundamental concept of the GCN is a layer-wise propagation rule, defined as:

\[
H^{(l+1)} = \sigma\left( \tilde{D}^{-\frac{1}{2}} \tilde{A} \tilde{D}^{-\frac{1}{2}} H^{(l)} W^{(l)} \right)
\]

where \( \tilde{A} = A + I_N \) is the adjacency matrix with added self-loops, \( \tilde{D} \) is the degree matrix of \( \tilde{A} \), \( H^{(l)} \) is the matrix of activation in the \( l \)-th layer, \( W^{(l)} \) is a layer-specific weight matrix, and \( \sigma \) is an activation function like ReLU. The final layer uses a softmax activation for classification:

\[
Z = \text{softmax} \left( \tilde{A} \text{ReLU} \left( \tilde{A} X W^{(0)} \right) W^{(1)} \right)
\]

This method efficiently integrates both node features and graph structure, and the adjacency matrix is normalized to prevent numerical instabilities and ensure effective gradient flow during training.

\subsubsection{GraphSAGE.} GraphSAGE is also an aggregation layer based on an inductive framework for generating node embeddings in large graphs \cite{hamilton2017inductive}. GraphSAGE learns embeddings by sampling and aggregating features from a node's local neighborhood, allowing it to generalize to previously unseen nodes. The three steps of GraphSAGE are described as follows: 1) sampling a fixed-size set of neighbors, 2) aggregating feature information using various functions (mean, LSTM, pooling), and 3) updating the node’s representation. Here are the typical aggregation functions used in GraphSAGE:
\begin{enumerate}
    \item \textbf{Mean Aggregator:}
    This function computes the mean of the feature vectors of the sampled neighbors.
    \[
    h_{\mathcal{N}(v)}^{k} = \frac{1}{|\mathcal{N}(v)|} \sum_{u \in \mathcal{N}(v)} h_u^{k-1}
    \]

    \item \textbf{LSTM Aggregator:}
    An LSTM is employed to aggregate the features from the neighborhood. This approach is order-sensitive.
    \[
    h_{\mathcal{N}(v)}^{k} = \text{LSTM}(\{h_u^{k-1} : u \in \mathcal{N}(v)\})
    \]
    Note: The LSTM aggregator processes the embeddings of the neighbors sequentially, and the output is the final hidden state of the LSTM.

    \item \textbf{Pooling Aggregator:}
    Applies a neural network followed by a max-pooling operation to the features of each neighbor.
    \[
    h_{\mathcal{N}(v)}^{k} = \max(\{\text{ReLU}(W \cdot h_u^{k-1} + b) : u \in \mathcal{N}(v)\})
    \]
\end{enumerate}

\subsubsection{Graph Attention Networks(GAT).}  Graph Attention Networks (GATs) is an aggregation layer with masked self-attention \cite{velickovic2017graph}. These self-attentional layers enable the GNN model to selectively focus on specific parts of a node's neighborhood, assigning different weights to neighboring nodes. This approach eliminates the need for computationally expensive matrix operations and does not require prior knowledge of the entire graph structure. The key components and equations of GATs are as follows:
\begin{enumerate}
    \item \textbf{Attention Coefficients:}
    This equation computes the attention coefficients that indicate the importance of each node's features to the central node. The attention coefficients are normalized using the softmax function to ensure they sum to one and can be interpreted as probabilities.
\\\[\alpha_{ij} = \frac{\exp\left(\text{LeakyReLU} \left(a^T [W h_i \| W h_j]\right)\right)}{\sum_{k \in N(i)} \exp\left(\text{LeakyReLU} \left(a^T [W h_i \| W h_k]\right)\right)}
\]
    \item \textbf{Aggregation:}
    The node features are aggregated using the weighted sum of the features of its neighbors, with weights given by the attention coefficients.
    \[
    \mathbf{h}_i' = \sigma \left( \sum_{j \in \mathcal{N}(i)} \alpha_{ij} \mathbf{W} \mathbf{h}_j \right)
    \]
\end{enumerate}

\subsection{Readout Layer}
The readout layer, also known as the graph-level pooling layer, plays a crucial role in the GNN model by transforming node-level embeddings into a graph-level representation. This representation is essential for tasks such as graph classification, graph regression, and entire graph embeddings. Conventionally, GNNs utilize simple pooling operations, such as averaging or summing, as described as follows:

\subsubsection{Mean Pooling Readout Layer} 
Mean pooling aggregates node features by computing the average. This process effectively captures the central tendency of feature distributions across the graph, which can be beneficial for representing the overall structure. The mean pooling operation can be mathematically described by:
\[
h_G = \frac{1}{|V|} \sum_{v \in V} h_v
\]
where \( h_v \) represents the node features, \( V \) is the set of all nodes in the graph, and \( h_G \) is the resulting graph-level representation.

\subsubsection{Add Readout Layer} 
Add pooling, or sum pooling, aggregates node features by computing the sum of all node embeddings. This method preserves the scale of node features, making it suitable for tasks where the magnitude of features impacts the output. The add pooling operation is expressed as:
\[
h_G = \sum_{v \in V} h_v
\]
where \( h_v \) and \( V \) are defined as above, and \( h_G \) is the graph-level representation.

While computationally efficient, these methods often limit capturing the intricate topology and heterogeneous relationships within the graph data.

\section{Methods}

\subsection {Data Description}
We collected a total of 383 Alzheimer's Disease patients, including 243 CN and 140 preclinical AD participants, from the Catholic Aging Brain Imaging(CABI) database. All participants, matched for age, gender, and education level, underwent T1-weighted MRI, resting-state functional MRI(rs-fMRI), and 18F-Florbetaben PET(FBB-PET). This database contains brain scans of patients who visited the outpatient clinic at the Catholic Brain Health Center, Yeouido St. Mary’s Hospital, The Catholic University of Korea, between 2017 and 2022. The study was conducted under ethical and safety guidelines set forth by the Institutional Review Board of Yeouido St. Mary’s Hospital, College of Medicine, The Catholic University of Korea (IRB number: SC22RIDI0153). The informed consent was waived by the IRB because we only used retrospective data. The detailed demographic information is described in Table~\ref{tab:sample_demographics}.

\begin{table}[h]
\centering
\caption{Demographic Information}
\label{tab:sample_demographics}
\begin{tabular}{@{}lc@{}}
\toprule
 & Mean $\pm$ standard deviation\\
\midrule
Age (years) & 70.96 $\pm$ 8.09\\
Education (years) & 11.56 $\pm$ 4.92\\
Gender (Male/female) & 114/270 \\
CDR (0/0.5/1.0) &  294/89/1 \\
PACC5 & 0.24 $\pm$ 0.59 \\
\bottomrule
\end{tabular}
\end{table}

\subsection {Data Preprocessing}
T1w MRI preprocessing was performed based on Micapipe to align normalized T1w MRI and parcellated brain into 374 regions, consisting of 360 cortical regions from Glasser atlas and 14 subcortical regions from aseg subcortical atlas \cite{glasser2016multi}. The rs-fMRI had been pre-processed using Micapipe \cite{cruces2022micapipe}, including the following steps: slice timing and head motion correction, skull stripping, intensity normalization, and band-pass filtering. The pre-processed data were registered onto the MNI152 standard space. The FBB-PET preprocessing was performed based on several steps: registration of the PET image to the subject’s T1-weighted MRI, partial volume correction using the PETPVC toolbox \cite{thomas2016petpvc}, inter-subject spatial normalization into MNI space. Finally, the standardized uptake value ratio(SUVR) was calculated for all parcellation by dividing the update of reference regions (i.e., cerebellumPons). 

\begin{figure}[h!]
\centering
    \includegraphics[width=\textwidth]{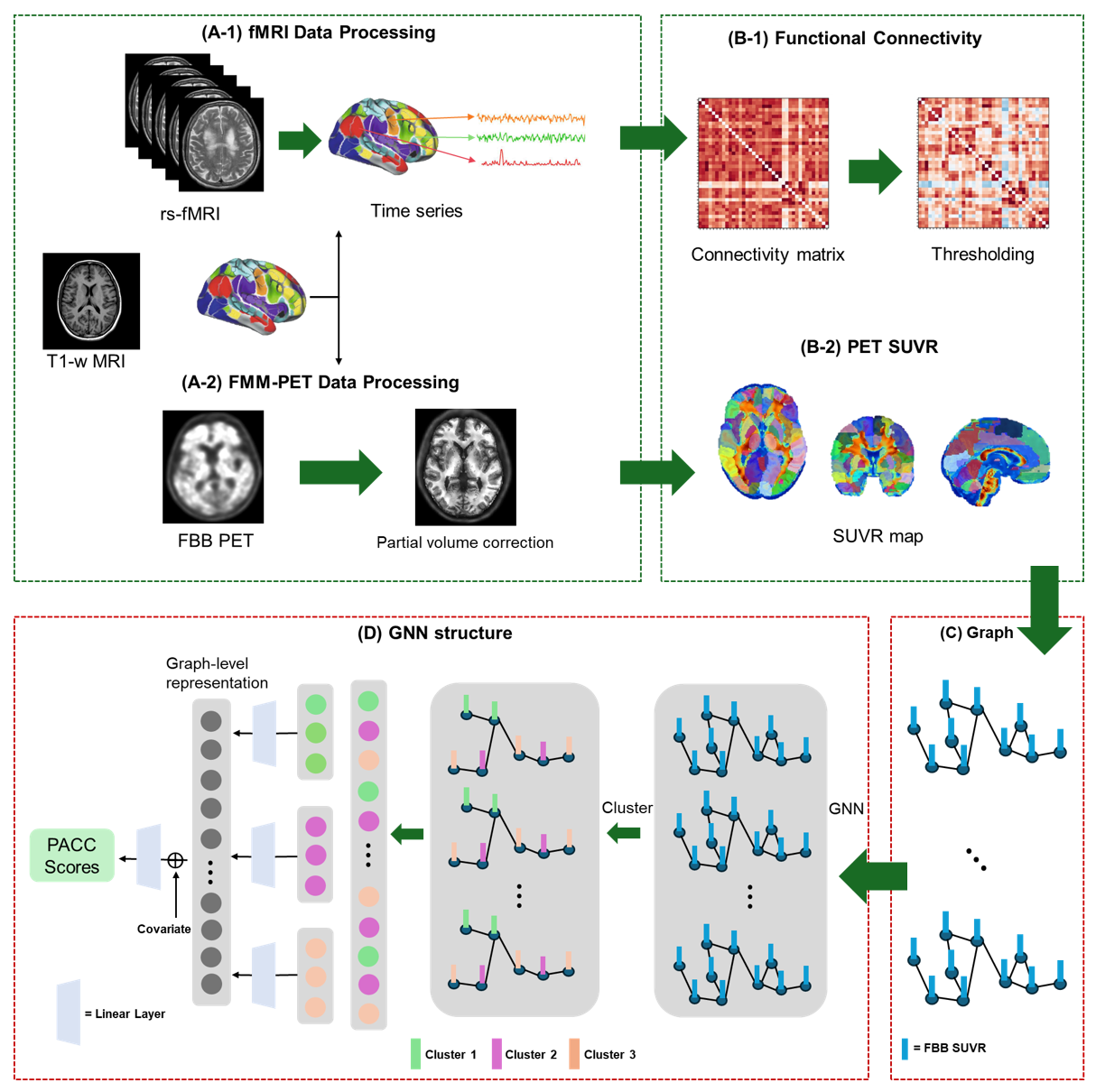}
    \caption{Overview of data processing and analysis pipeline for neuroimaging and graph neural network application. (A-1) fMRI data processing involves acquiring fMRI time-series, followed by brain parcellation and extraction of time courses. (A-2) For FBB-PET data, The T1-weighted images are processed by partial volume correction to produce PET SUVR maps. (B-1) Functional connectivity is calculated from fMRI data, resulting in connectivity matrices that are further refined through thresholding. (B-2) PET SUVR maps are obtained from processed FBB-PET images. (C) The processed data are then represented as graphs, which are input into a GNN. (D) The GNN structure processes these graphs to yield graph-level representations and predicts clinical scores such as PACC. } 
    \label{fig:architecture}
\end{figure}

\subsection {Brain-aware Graph Neural Network}

\subsubsection{Motivation} Unlike social networks or chemical molecular networks, brain networks have the unique characteristic that nodes(i.e., brain regions) are organized into networks based on similar roles or functions. Therefore, simple readout layers that average or sum node features are insufficient for capturing the complex and heterogeneous characteristics of brain networks. To overcome limitations, we propose a novel readout layer clustering brain regions regarding similar tasks and allowing the GNN model to represent details brain characteristics. By leveraging these clusters, we aim the proposed GNN model to learn the intricate and heterogeneous features of brain networks, leading to more accurate and meaningful graph-level representations. 

\subsubsection{Brain-aware Readout Layer (BA readout layer)} The BA readout layer is designed to effectively capture the complex and heterogeneous characteristics of brain networks by clustering brain regions based on functional connectivity and node embedding. As shown in Figure~\ref{fig:architecture} (D), the BA readout layer is expressed as:

\[h_{BA} = [f(h_{c_1}), f(h_{c_2}), ..., f(h_{c_i})]^T, 
\text{where }   h_{c_i} = \frac{1}{|V_{c_i}|} \sum_{v \in V_{c_i}} h_v\]
where \(h_{BA}\) represents group-level representation, \(f(.)\) denotes a linear transformation function used to aggregate embeddings, \(V_{c_i}\) represents a set of nodes in $c_i$ cluster, and \(h_{c_i}\) denotes a cluster-level embedding in the $c_i$ cluster. 

Briefly, the brain-aware clusters and their members have been selected by using a simple auxiliary neural network. An auxiliary network consisted of a single-layer artificial neural network with softmax function feed position embedding of each node, \(\mathbf{C} = \text{softmax}(f(\mathcal{P}))\), and compute its probability of belonging to each cluster. The position embedding,  \(\mathcal{P} = [p_{1}, p_{2}, ..., p_{i}]\in \mathbb{R}^{N \times N}\), is a one-hot vector, where the $i$-th entry of \( p_{i} \) is 1 and all other entries are 0, which make embedding invariant to the order of node. 

\[\mathbf{C} = [c_1, c_2, ..., c_i] \in \mathbb{R}^{N \times p}\]

where $N$ denotes number of nodes in the graph, $p$ denotes  number of brain-aware cluster, and \(c_i\) denotes selection probability vector of the $i$-th cluster for all nodes.

The cluster-level embedding, \(f(h_{c_i})\), is computed by aggregating node embedding, $ h_{c_i}$, based on $c_i$, selected members of the $c_i$ cluster. 
Finally, the group-level representation is obtained by concatenating all cluster-level embeddings. This representation is then used in further analysis.

\subsubsection{Architecture} The pre-processed fMRI and FBB-PET are utilized to construct the input graph, as described in Figure~\ref{fig:architecture} (A)-(C). The proposed network model encodes node-level representation based on the existing GNN encoder and then graph-level representation based on the proposed BA readout layer. Using the BA readout layer, nodes are clustered into groups based on their similarities, indicated by different node colors in Figure~\ref{fig:architecture} (D), and are aggregated to generate a comprehensive graph-level representation that captures the intricate structure of the brain network.  The final representation is subsequently concatenated with covariate variables (e.g., age, gender, and education) and passed through a linear layer to predict the PACC score.  

{\itshape Step 1: Graph Construction.} 
For given neuroimage, the input graph \( g \) is represented as a set of triplet \(\{(\mathcal{V}, \mathbf{E},\mathcal{P})\}\). \( \mathcal{V} \) represents the set of nodes, expressed as \( \mathcal{V} = \{ v_1, v_2, \ldots, v_N \} \in \mathbb{R}^{N \times d_{in}} \), where each \( v_i \) can be neuroimaging measures with  $d_{in}$ features, such as brain region volume, SUVR of PET image, or thickness. \( \mathbf{E} \) is the set of edges, consisting of pairs \( (e_i, e_j) \). Each \( (e_i, e_j) \) denotes the connectivity information between nodes (i.e., brain regions), and represents a non-negative undirected connection (either functional or structural connectivity).  \(\mathcal{P}\) represents the positional information of the nodes, where the $i$-th entry of \( p_{i} \) is 1 and all other entries are 0. The node position is assigned based on the brain atlas so that the nodes of all graph instances are ordered in a consistent sequence. Additionally, \( y =\{ y_1, y_2, \ldots, y_{N} \}\) represents the label for graph classification or the value for regression tasks. 

{\itshape Step 2: Node Embedding.} 
In this step, we perform node embedding using the GNN models mentioned in Section 2.1. The GNN model, such as GCN, GraphSAGE, or GAT, aggregates node attributes to generate node-wise embedding, \(h_v\). \begin{align}
h_v^{(k)} &= \text{COMBINE}^{(k)} \left( \text{AGGREGATE}^{(k)} \left( \{ h_u^{(k-1)} : u \in \mathcal{N}(v) \} \right), h_v^{(k)} \right)
\end{align}
Here, \( h_v^{(k)} \in \mathbb{R}^{N \times d} \) represents the embedding of node \( v \) at layer \( k \), where \( d \) is the hidden dimension. 
\(\text{AGGREGATE}^{(k)}\) denotes the aggregation function at layer \( k \), and \(\text{COMBINE}^{(k)}\) represents the combination function at layer \( k \). 
The function AGGREGATE processes local feature information from the node's neighbors to capture the local structural context within the graph. The $\text{COMBINE}$ function integrates the aggregated neighbor information with the node's existing features to update and refine its embedding. 
%The Ba-Readout Layer is designed to be compatible with various GNN architectures by fitting into this general framework.

{\itshape Step 3: Graph-level representation.} 
Finally, we applied the BA readout layer to generate the graph-level representation for each input graph. Briefly, the brain-aware clusters and their members have been selected by using a simple auxiliary neural network, and the graph-level representation is obtained by concatenating cluster-level embeddings. 

\section{Experiments and results} 
\subsection{Experimental Setups}
We have applied the proposed BA readout layer on various GNN models and assessed their predictive performance compared with the model with conventional readout layers. Many researchers have successfully adopted the GNN model for brain connectivity analysis. We have carefully chosen three GNN models: 1) GCN, 2) GraphSAGE, and 3) GAT, and two readout layers: 1) mean-pooling layer, and 2) add pooling layer. For all benchmark algorithms, we have constructed a graph where a cortico-cortical functional connectivity is based on Pearson's correlation with a predefined atlas as the edges of the graph and the pre-calculated SUVR for each brain region as node attributes of the graph.

In this study, we focus on the early detection of AD at the preclinical stage, which predicting the PACC score. We have applied a five-fold cross-validation strategy to examine the performance of the model, in terms of $R^{2}$  score.  The model hyperparameters, such as hidden dimensions and learning rates, optimizing the number of clusters for each model, are tuned by Bayesian search on the training set implemented in Sweep methodology from Weights and Biases (WandB) \cite{ferrell2023fine}. Bayesian search strategy efficiently searches the most effective combination of hyperparameters using Bayesian optimization. Once optimal hyper-parameters are determined, the trained model is applied to the test set to generate the final performance.

\subsection {Performance Comparison Across Readout Layer on Various GNN Model}
To assess the effectiveness of the proposed method(i.e., BA readout layer), we have compared the predictive power of various GNN models with conventional readout layer, such as mean readout and add readout layer, against those using BA readout layer. The predictive power of the model is evaluated by predicting the PACC score of CN and preclinical AD subjects.  

\begin{table}[b]
\centering
\caption{Comparison of Readout Methods Across Different GNN Architectures with Performance Measured by $R^{2}$ Score}
\label{tab:readout_comparison}
\begin{tabular}{@{}lcccc@{}}
\toprule
& \textbf{Mean}& \textbf{Add}&  \textbf{Yeo BA readout}&  \textbf{adaptive BA readout}\\ \midrule
\textbf{GCN}& $0.1539 \pm 0.04$& $0.0877 \pm 0.07$& $\mathbf{ 0.1734 \pm 0.05}$& $ 0.1517 \pm 0.02$\\
\textbf{GraphSAGE}& $0.1917 \pm 0.02$& $0.0441 \pm 0.05$& $0.1921\pm 0.03$& $\mathbf{0.1969 \pm 0.02}$\\
\textbf{GAT}& $0.2019 \pm 0.03$& $0.0758 \pm 0.03$& $\mathbf{0.2141 \pm 0.08}$& $0.2044 \pm 0.03$\\ \bottomrule
\end{tabular}
\end{table}

Table ~\ref{tab:readout_comparison} indicated that our model outperforms to predict PACC for all GNN models compared with those with mean and add readout layer. Specifically, the GCN and GAT with BA readout layer using Yeo network yield the best performance, $0.1734 \pm 0.05$, and $ 0.2141 \pm 0.08$, of $R^{2}$ scores, respectively, and the GraphSAGE with adaptive BA readout layer yield the best performance, $0.1969 \pm 0.02$, of $R^{2}$ scores. 

\subsection {Effectiveness of Prior Knowledge in BA Readout Layer}
Additionally, we examine the effectiveness of incorporating prior knowledge into the BA readout layer. In detail, we apply the well-known Yeo 7 functional network as a predefined cluster of BA readout layer, called as Yeo BA readout layer, and compared with the proposed BA readout layer, called adaptive BA readout layer \cite{yeo2011organization}.  Overall, the integration of prior knowledge with the proposed layer (i.e., Yeo BA readout layer) demonstrates enhanced performance compared to layers without such knowledge (i.e., adaptive BA readout layer). However, we noted that the adaptive BA readout layer exhibited a lower standard deviation than the Yeo BA readout layer. This suggests that the adaptive BA readout layer offers greater robustness and stability in its performance.

\begin{figure}[t!]
\centering
    \includegraphics[width=\textwidth]{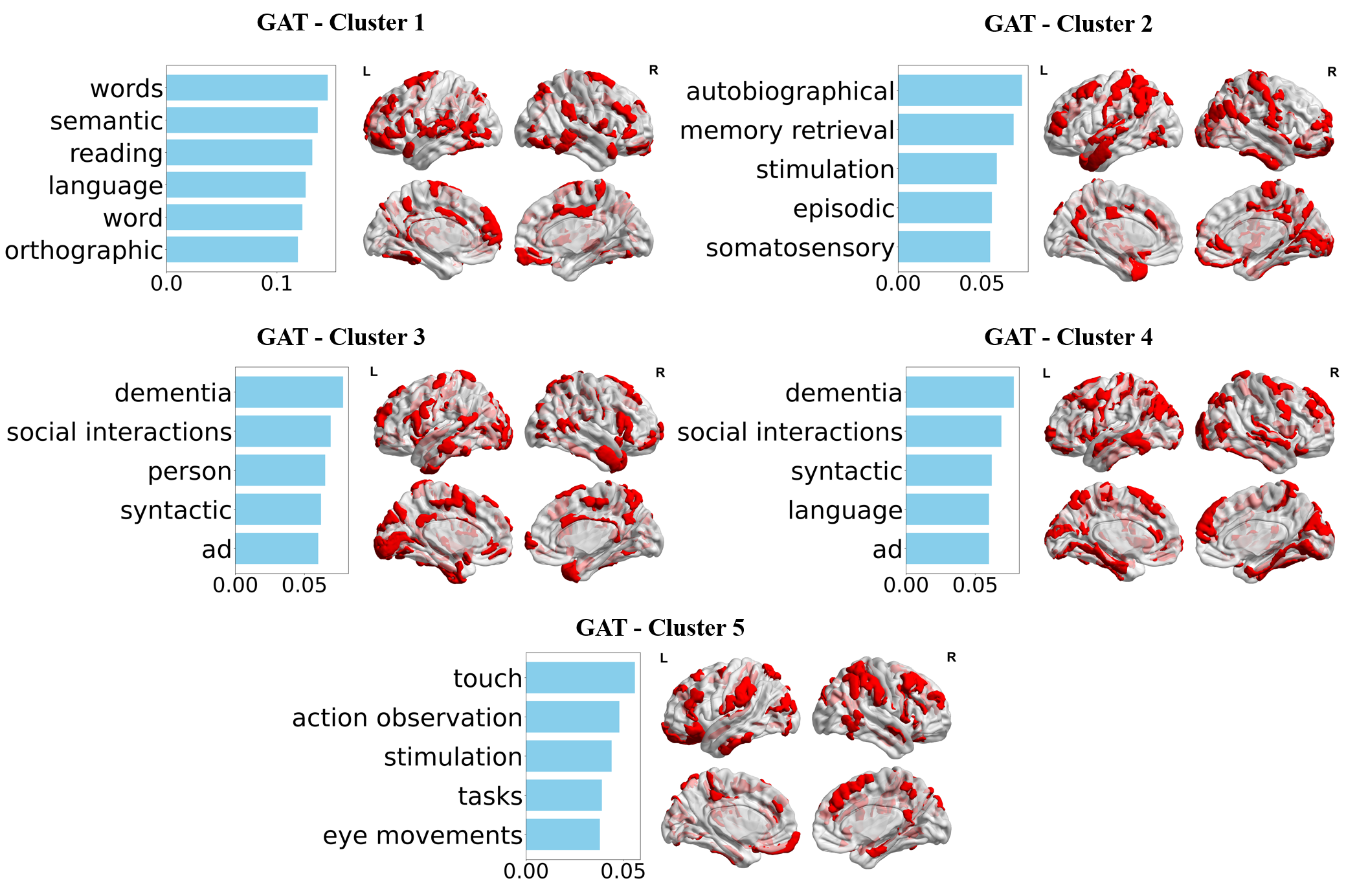}
    \caption{Visualization of brain function clusters derived from GAT} 
    \label{fig:GAT Cluster}
\end{figure}

\subsection {Interpretability of the GNN Model with BA Readout Layer}
The advantage of the adaptive BA readout layer is that be able to select a set of task-specific brain regions of the clusters, regardless of the GNN model. Particularly, the members of clusters are AD-specific brain regions, which play roles in memory, cognitive, or sensory function, that improve the model’s interpretability. We utilized Neurosynth \cite{yarkoni2011large}, a platform that synthesizes large amounts of human brain imaging data to decode patterns of neural activity associated with psychological terms and cognitive states to interpret selected brain regions of our model. Top 5 Neurosynth topics were investigated and the topics related to anatomical terminology were excluded. 

Figure~\ref{fig:GAT Cluster} illustrates five distinct clusters and their top five Neurosynth topics identified by the GAT with the BA readout layer. The brain regions within Cluster 1 are associated with language-related tasks, those within Cluster 2 are linked to memory functions, those within Clusters 3 and 4 are associated with dementia and social interactions, and those within Cluster 5 are related to sensory-motor functions. We found that the model identified the anterior cingulate, temporal pole, anterior insula, orbitofrontal cortex, and prefrontal cortex as members of clusters 3 and 4. These regions play important roles in cognitive functions and emotional responses or memory decline which is associated with AD  \cite{molnar2022anterior,carter1999contribution,arnold1994neuropathologic}.

\begin{figure}[t!]
\centering
    \includegraphics[width=\textwidth]{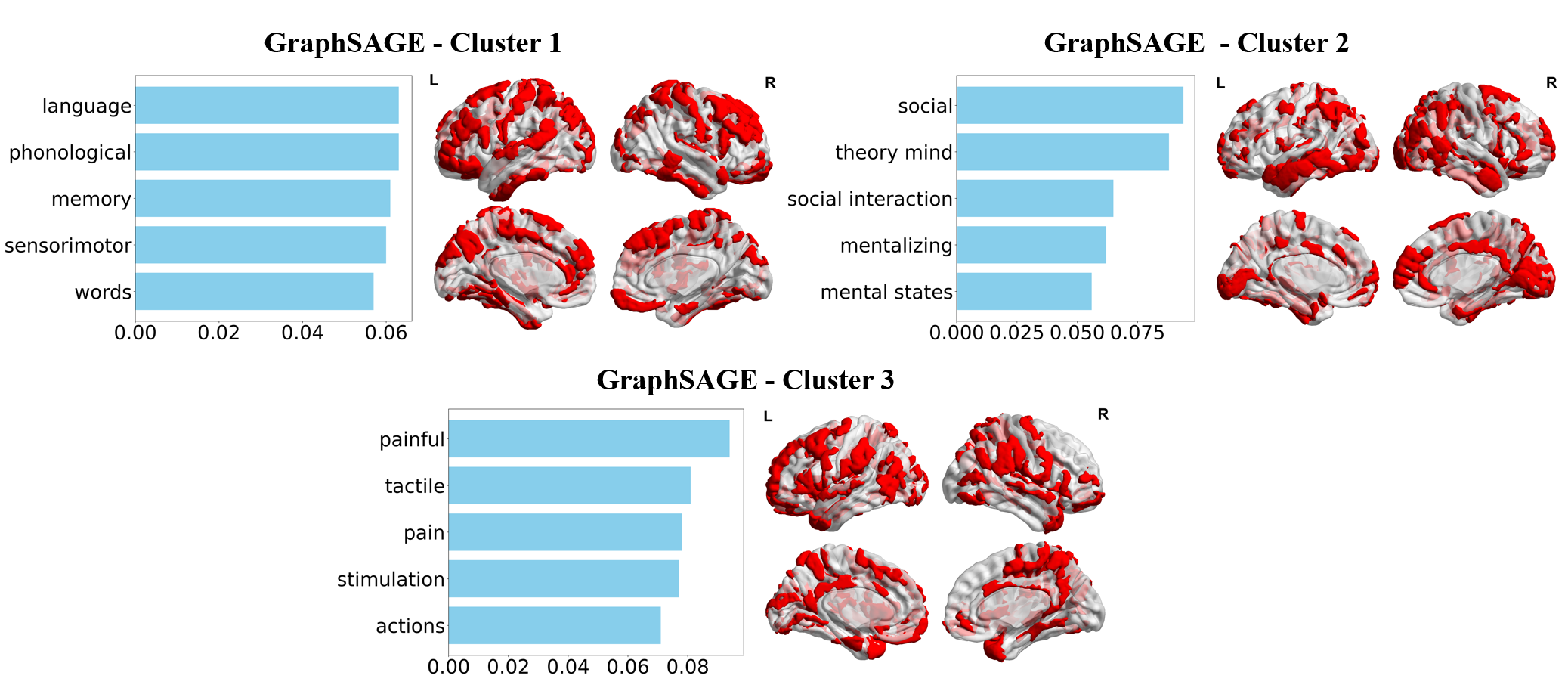}
    \caption{Visualization of brain function clusters derived from GraphSAGE} 
    \label{fig:Sage cluster}
\end{figure}

\begin{figure}[h!]
\centering
    \includegraphics[width=\textwidth]{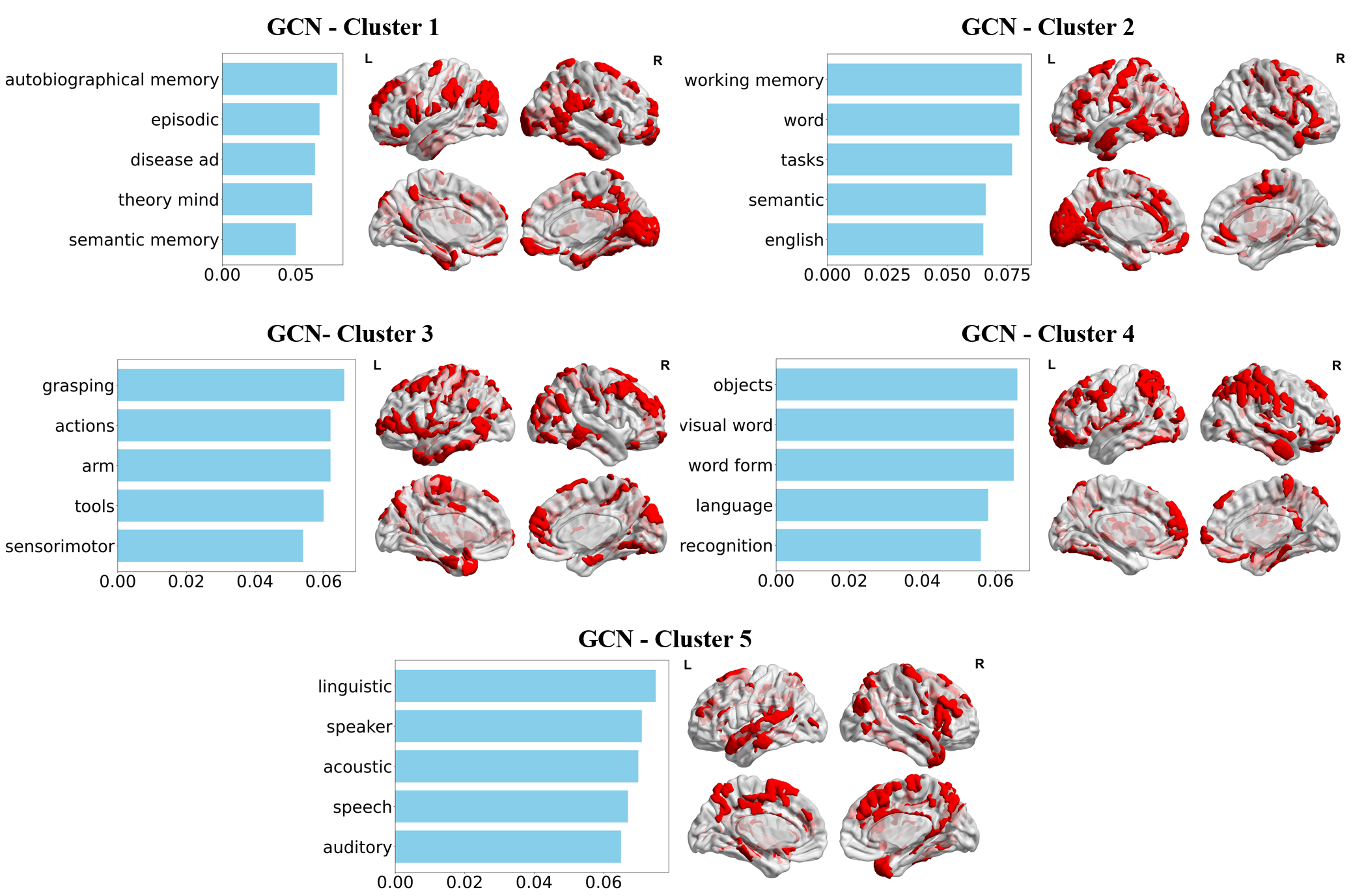}
    \caption{Visualization of brain function clusters derived from GCN} 
    \label{fig:GCN cluster}
\end{figure}

Figure~\ref{fig:Sage cluster} illustrates three distinct clusters and their top five Neurosynth topics identified by the GraphSAGE with the BA readout layer. The brain regions within cluster 1 are associated with language and memory processing, which is crucial for detecting early signs of cognitive decline related to linguistic abilities and memory retention \cite{clarke2021comparison,beltrami2018speech}. The brain regions within cluster 2 are linked to social cognition, essential for understanding and interacting with others, which can be impaired in Alzheimer’s disease \cite{demichelis2020empathy,wilson2024disrupted}. Those within Cluster 3 are associated with sensory processing and pain. 

Figure~\ref{fig:GCN cluster} illustrates five distinct clusters and their top five Neurosynth topics identified by the GCN with the BA readout layer. The brain regions within each cluster are associated with specific brain functions relevant to cognitive disorders, particularly Alzheimer's disease. For example, the brain regions within clusters 1 and 2 are associated with memory \cite{verma2012semantic}, those within clusters 4 and 5 are linked to language tasks, essential for communication, and which deteriorate over the course of Alzheimer's disease \cite{swords2018auditory,massoud2002word}.

\section {Conclusion}

\noindent AD presents significant challenges in both diagnosis and treatment, particularly due to its heterogeneous manifestation and the varying progression among individuals. This study introduces a novel approach to improve the interpretability and predictive power of GNNs in analyzing brain networks for early diagnosis of AD. We propose a brain-aware readout layer (BA readout layer) that clusters brain regions based on functional connectivity and node embedding, allowing the GNN model to capture intricate and heterogeneous characteristics of brain networks more effectively.

Our experiments demonstrate that GNN models incorporating the BA readout layer outperform those using conventional readout layers in predicting the PACC score. Specifically, the GCN and GAT models with the Yeo BA readout layer, as well as the GraphSAGE model with the adaptive BA readout layer, achieve the highest $R^2$ scores, indicating superior predictive performance.

Moreover, the adaptive BA readout layer provides robustness and stability, exhibiting lower standard deviation compared to the Yeo BA readout layer. Furthermore, unlike the Yeo BA readout, which focuses solely on functional connectivity, the adaptive BA readout considers functional connectivity as well as node embedding, allowing for a richer and more meaningful graph-level representation. This suggests that the adaptive approach is more reliable across different datasets and scenarios. The interpretability of our model is further enhanced by the ability to identify task-specific brain regions within the clusters, which are associated with key cognitive functions and areas known to be affected by AD.

In summary, our proposed brain-aware readout layer for GNNs offers a promising tool for improving the diagnosis of Alzheimer's disease by leveraging the complex patterns inherent in brain data. By effectively clustering brain regions based on functional connectivity and node embedding, our approach enhances both the interpretability and performance of GNN models in neuroimaging analysis. Future research can further explore the potential of this method in other neurological disorders and refine the clustering techniques to capture even more nuanced brain network characteristics.

\begin{credits}
\subsubsection{\ackname} This work was partly supported by Institute of Information \& communications Technology Planning \& Evaluation (IITP) grant funded by the Korea government(MSIT) (No.RS-2021-II212068, Artificial Intelligence Innovation Hub, No.2019-0-01842, Artificial Intelligence Graduate School Program (GIST)). This work was also partly supported by the National Research Foundation of Korea (NRF) grant funded by the Korean government (Ministry of Science and ICT) (NRF-2022R1F1A106 8529, 2022R1I1A1A01053710). We also appreciate the high-performance GPU computing support of HPC-AI Open Infrastructure via GIST SCENT.

\subsubsection{\discintname}
The authors have no competing interests to declare that are
relevant to the content of this article.
\end{credits}
\bibliographystyle{splncs04}
\bibliography{ref}

\begin{thebibliography}{10}
\providecommand{\url}[1]{\texttt{#1}}
\providecommand{\urlprefix}{URL }
\providecommand{\doi}[1]{https://doi.org/#1}

\bibitem{silverman1999clinical}
Silverman, D.H., Small, G.W., Phelps, M.E.: Clinical value of neuroimaging in the diagnosis of dementia: sensitivity and specificity of regional cerebral metabolic and other parameters for early identification of alzheimer's disease. Clinical Positron Imaging  \textbf{2}(3),  119--130 (1999)

\bibitem{serrano2021apoe}
Serrano-Pozo, A., Das, S., Hyman, B.T.: Apoe and alzheimer's disease: advances in genetics, pathophysiology, and therapeutic approaches. The Lancet Neurology  \textbf{20}(1),  68--80 (2021)

\bibitem{selkoe2016amyloid}
Selkoe, D.J., Hardy, J.: The amyloid hypothesis of alzheimer's disease at 25 years. EMBO molecular medicine  \textbf{8}(6),  595--608 (2016)

\bibitem{ael2003early}
A{\"e}l~Chetelat, G., Baron, J.C.: Early diagnosis of alzheimer’s disease: contribution of structural neuroimaging. Neuroimage  \textbf{18}(2),  525--541 (2003)

\bibitem{hanseeuw2019association}
Hanseeuw, B.J., Betensky, R.A., Jacobs, H.I., Schultz, A.P., Sepulcre, J., Becker, J.A., Cosio, D.M.O., Farrell, M., Quiroz, Y.T., Mormino, E.C., et~al.: Association of amyloid and tau with cognition in preclinical alzheimer disease: a longitudinal study. JAMA neurology  \textbf{76}(8),  915--924 (2019)

\bibitem{donohue2014preclinical}
Donohue, M.C., Sperling, R.A., Salmon, D.P., Rentz, D.M., Raman, R., Thomas, R.G., Weiner, M., Aisen, P.S., et~al.: The preclinical alzheimer cognitive composite: measuring amyloid-related decline. JAMA neurology  \textbf{71}(8),  961--970 (2014)

\bibitem{jack2018nia}
Jack~Jr, C.R., Bennett, D.A., Blennow, K., Carrillo, M.C., Dunn, B., Haeberlein, S.B., Holtzman, D.M., Jagust, W., Jessen, F., Karlawish, J., et~al.: Nia-aa research framework: toward a biological definition of alzheimer's disease. Alzheimer's \& Dementia  \textbf{14}(4),  535--562 (2018)

\bibitem{salvatore2015magnetic}
Salvatore, C., Cerasa, A., Battista, P., Gilardi, M.C., Quattrone, A., Castiglioni, I., Initiative, A.D.N.: Magnetic resonance imaging biomarkers for the early diagnosis of alzheimer's disease: a machine learning approach. Frontiers in neuroscience  \textbf{9}, ~307 (2015)

\bibitem{hoilund2023fdg}
H{\o}ilund-Carlsen, P.F., Revheim, M.E., Costa, T., Kepp, K.P., Castellani, R.J., Perry, G., Alavi, A., Barrio, J.R.: Fdg-pet versus amyloid-pet imaging for diagnosis and response evaluation in alzheimer’s disease: Benefits and pitfalls. Diagnostics  \textbf{13}(13), ~2254 (2023)

\bibitem{chapleau2022role}
Chapleau, M., Iaccarino, L., Soleimani-Meigooni, D., Rabinovici, G.D.: The role of amyloid pet in imaging neurodegenerative disorders: a review. Journal of Nuclear Medicine  \textbf{63}(Supplement 1),  13S--19S (2022)

\bibitem{sheline2013resting}
Sheline, Y.I., Raichle, M.E.: Resting state functional connectivity in preclinical alzheimer’s disease. Biological psychiatry  \textbf{74}(5),  340--347 (2013)

\bibitem{wu2020comprehensive}
Wu, Z., Pan, S., Chen, F., Long, G., Zhang, C., Philip, S.Y.: A comprehensive survey on graph neural networks. IEEE transactions on neural networks and learning systems  \textbf{32}(1),  4--24 (2020)

\bibitem{boll2024graph}
Boll, H.O., Amirahmadi, A., Ghazani, M.M., de~Morais, W.O., de~Freitas, E.P., Soliman, A., Etminani, K., Byttner, S., Recamonde-Mendoza, M.: Graph neural networks for clinical risk prediction based on electronic health records: A survey. Journal of Biomedical Informatics p. 104616 (2024)

\bibitem{zhang2023graph}
Zhang, L., Zhao, Y., Che, T., Li, S., Wang, X.: Graph neural networks for image-guided disease diagnosis: A review. iRADIOLOGY  \textbf{1}(2),  151--166 (2023)

\bibitem{zhou2020graph}
Zhou, J., Cui, G., Hu, S., Zhang, Z., Yang, C., Liu, Z., Wang, L., Li, C., Sun, M.: Graph neural networks: A review of methods and applications. AI open  \textbf{1},  57--81 (2020)

\bibitem{jiang2021could}
Jiang, D., Wu, Z., Hsieh, C.Y., Chen, G., Liao, B., Wang, Z., Shen, C., Cao, D., Wu, J., Hou, T.: Could graph neural networks learn better molecular representation for drug discovery? a comparison study of descriptor-based and graph-based models. Journal of cheminformatics  \textbf{13},  1--23 (2021)

\bibitem{li2021braingnn}
Li, X., Zhou, Y., Dvornek, N., Zhang, M., Gao, S., Zhuang, J., Scheinost, D., Staib, L.H., Ventola, P., Duncan, J.S.: Braingnn: Interpretable brain graph neural network for fmri analysis. Medical Image Analysis  \textbf{74},  102233 (2021)

\bibitem{kipf2016semi}
Kipf, T.N., Welling, M.: Semi-supervised classification with graph convolutional networks. arXiv preprint arXiv:1609.02907  (2016)

\bibitem{hamilton2017inductive}
Hamilton, W., Ying, Z., Leskovec, J.: Inductive representation learning on large graphs. Advances in neural information processing systems  \textbf{30} (2017)

\bibitem{velickovic2017graph}
Velickovic, P., Cucurull, G., Casanova, A., Romero, A., Lio, P., Bengio, Y., et~al.: Graph attention networks. stat  \textbf{1050}(20),  10--48550 (2017)

\bibitem{glasser2016multi}
Glasser, M.F., Coalson, T.S., Robinson, E.C., Hacker, C.D., Harwell, J., Yacoub, E., Ugurbil, K., Andersson, J., Beckmann, C.F., Jenkinson, M., et~al.: A multi-modal parcellation of human cerebral cortex. Nature  \textbf{536}(7615),  171--178 (2016)

\bibitem{cruces2022micapipe}
Cruces, R.R., Royer, J., Herholz, P., Larivi{\`e}re, S., De~Wael, R.V., Paquola, C., Benkarim, O., Park, B.y., Degr{\'e}-Pelletier, J., Nelson, M.C., et~al.: Micapipe: A pipeline for multimodal neuroimaging and connectome analysis. NeuroImage  \textbf{263},  119612 (2022)

\bibitem{thomas2016petpvc}
Thomas, B.A., Cuplov, V., Bousse, A., Mendes, A., Thielemans, K., Hutton, B.F., Erlandsson, K.: Petpvc: a toolbox for performing partial volume correction techniques in positron emission tomography. Physics in Medicine \& Biology  \textbf{61}(22), ~7975 (2016)

\bibitem{ferrell2023fine}
Ferrell, B.J.: Fine-tuning strategies for classifying community-engaged research studies using transformer-based models: algorithm development and improvement study. JMIR Formative Research  \textbf{7},  e41137 (2023)

\bibitem{yeo2011organization}
Yeo, B.T., Krienen, F.M., Sepulcre, J., Sabuncu, M.R., Lashkari, D., Hollinshead, M., Roffman, J.L., Smoller, J.W., Z{\"o}llei, L., Polimeni, J.R., et~al.: The organization of the human cerebral cortex estimated by intrinsic functional connectivity. Journal of neurophysiology  (2011)

\bibitem{yarkoni2011large}
Yarkoni, T., Poldrack, R.A., Nichols, T.E., Van~Essen, D.C., Wager, T.D.: Large-scale automated synthesis of human functional neuroimaging data. Nature methods  \textbf{8}(8),  665--670 (2011)

\bibitem{molnar2022anterior}
Molnar-Szakacs, I., Uddin, L.Q.: Anterior insula as a gatekeeper of executive control. Neuroscience \& Biobehavioral Reviews  \textbf{139},  104736 (2022)

\bibitem{carter1999contribution}
Carter, C.S., Botvinick, M.M., Cohen, J.D.: The contribution of the anterior cingulate cortex to executive processes in cognition. Reviews in the Neurosciences  \textbf{10}(1),  49--58 (1999)

\bibitem{arnold1994neuropathologic}
Arnold, S.E., Hyman, B.T., Van~Hoesen, G.W.: Neuropathologic changes of the temporal pole in alzheimer's disease and pick's disease. Archives of neurology  \textbf{51}(2),  145--150 (1994)

\bibitem{clarke2021comparison}
Clarke, N., Barrick, T.R., Garrard, P.: A comparison of connected speech tasks for detecting early alzheimer’s disease and mild cognitive impairment using natural language processing and machine learning. Frontiers in Computer Science  \textbf{3},  634360 (2021)

\bibitem{beltrami2018speech}
Beltrami, D., Gagliardi, G., Rossini~Favretti, R., Ghidoni, E., Tamburini, F., Calz{\`a}, L.: Speech analysis by natural language processing techniques: a possible tool for very early detection of cognitive decline? Frontiers in aging neuroscience  \textbf{10}, ~369 (2018)

\bibitem{demichelis2020empathy}
Demichelis, O.P., Coundouris, S.P., Grainger, S.A., Henry, J.D.: Empathy and theory of mind in alzheimer’s disease: A meta-analysis. Journal of the International Neuropsychological Society  \textbf{26}(10),  963--977 (2020)

\bibitem{wilson2024disrupted}
Wilson, N.A., Ahmed, R., Piguet, O., Irish, M.: Disrupted social perception in frontotemporal dementia and alzheimer's disease--associated cognitive processes and clinical implications. Journal of the Neurological Sciences  \textbf{458},  122902 (2024)

\bibitem{verma2012semantic}
Verma, M., Howard, R.J.: Semantic memory and language dysfunction in early alzheimer's disease: a review. International journal of geriatric psychiatry  \textbf{27}(12),  1209--1217 (2012)

\bibitem{swords2018auditory}
Swords, G.M., Nguyen, L.T., Mudar, R.A., Llano, D.A.: Auditory system dysfunction in alzheimer disease and its prodromal states: A review. Ageing research reviews  \textbf{44},  49--59 (2018)

\bibitem{massoud2002word}
Massoud, F., Chertkow, H., Whitehead, V., Overbury, O., Bergman, H.: Word-reading thresholds in alzheimer disease and mild memory loss: a pilot study. Alzheimer Disease \& Associated Disorders  \textbf{16}(1),  31--39 (2002)

\end{thebibliography}

\end{document}